\documentstyle[prl,aps,multicol,epsfig]{revtex}
\renewcommand{\narrowtext}
\renewcommand{\widetext}

\begin{document}
\title{Optimized design of universal two-qubit gates}
\author{I. A. Grigorenko and D. V. Khveshchenko}
\address{Department of Physics and Astronomy, University of North Carolina,
Chapel Hill, NC 27599}
\maketitle

\begin{abstract}
We construct optimized implementations of the CNOT 
and other universal two-qubit gates that, unlike many of
the previously proposed protocols, are carried out in a single step.
The new protocols require tunable inter-qubit couplings but, 
in return, show a significant
improvements in the quality of gate operations. 
Our optimization procedure can be further extended to the 
combinations of elementary two-qubit as well as irreducible 
many-qubit gates.
\end{abstract}

According to one of the central results obtained in 
quantum information theory, an arbitrarily complex
quantum protocol can be decomposed into a sequence of single-qubit rotations and
two-qubit gates \cite{DiVincenzo}. However, despite its providing a convenient means of
designing logical circuits, in practice such a decomposition may 
not necessarily achieve the shortest possible times of operation and, consequently, 
the lowest possible decoherence rates. In any practical implementations, however, the latter are
crucially important, for any realistic qubit system would always
suffer a detrimental effect of its dissipative environment.

Recently, there have been various attempts to improve the
performance of universal quantum gates by searching for their
optimal implementations among the entire class of two-qubit Hamiltonians 
with the most general time dependent coefficients. 
However, a typical outcome of such a tour-de-force 
variational search \cite{salomaa}
tends to be a complicated sequence of highly irregular pulses
whose physical content often remains largely obscure.

In the search of a more sophisticated analytical approach, a number of authors invoked
optimal control theory with the goal of implementing a desired
unitary transformation independently of the initial state. The
resulting complex system of nonlinear integral-differential
equations can be solved numerically with the help of the Krotov or similar iterative
algorithms \cite{kosloff}.

Conceivably, a significantly simpler alternative to the above
approaches would be a straightforward implementation of a given unitary
transformation in the smallest possible number of steps, during
each of which the Hamiltonian remains constant. 
A well known example of this kind is provided by the two-qubit SWAP gate 
which can be readily implemented (up to a global phase)
with the use of the spin-rotationally invariant Heisenberg inter-qubit coupling that
remains constant during the gate operation.

In this paper, we construct one-step
implementations of some widely used universal gates. In contrast to the
previous works where a constant decoherence rate was assumed and, therefore,
the overall loss of coherence accumulated during a
gate operation would be evaluated solely on the basis of its total duration, 
we quantify the adverse effect of the environment by actually
solving the corresponding master equation for the density matrix of 
the coupled qubits. In this way, we account for the fact 
that the decoherence rates generally depend upon (and vary with) 
the adjustable parameters of the Hamiltonian.

The problem of implementing a given unitary transformation
in the course of quantum mechanical evolution of a generic
two-qubit system can be formulated as the condition of a minimum deviation
\begin{equation}
||\hat{X}-{\hat T}\exp(-i\int_0^{t_{0}}{\hat{{H}}(x_i(t'))dt'})||\to min
\end{equation}
between 
the time-ordered evolution operator governed by the 
Hamiltonian $\hat{H}$ and the target unitary transformation $\hat{X}$. 
Here $x_i(t)$ represent the control 
parameters whose physically attainable values
are generally bounded ($|x_i(t)|<a_i$), and the deviation measure is given 
by the Frobenius trace norm $||\hat{Y}||=\sqrt{Tr [\hat{Y}^{\dagger}\hat{Y} ]}$.

In the case of interest, the Hamiltonian 
\begin{equation}
\label{4}
{\hat{H}}=\sum_{i=1,2}{\hat{\vec \sigma}}_{i}
({\vec B}_i(t)+{\vec h}_i(t))
+\sum_{a=x,y,z}J_{a}(t){\hat{\sigma}}^a_{1}{\hat{\sigma}}^a_{2}
\end{equation}
operates on a direct product of two (pseudo)spin Hilbert spaces. 

The tunable local fields ${\vec B}_i=(\Delta_i,0,\epsilon_i)$ encode 
the tunneling amplitudes $\Delta_{1,2}(t)$ and biases $\epsilon_{1,2}(t)$  
of the individual qubits, while their random 
(and uncorrelated, $<h_1(t)h_2(t^\prime)>=0$) counterparts ${\vec h}_i=(0,0,h_i)$ 
represent the noisy environment in the worst-case scenario of two independent dissipative reservoirs. 
As compared to the opposite limit of two strongly correlated reservoirs (collective decoherence), 
in this case no decoherence-free subspace can possibly exist.

In the standard basis where $\sigma^z_{1,2}$ are diagonal, 
the noiseless part of Eq.(\ref{4}) takes the form
\begin{equation} 
\label{5}
\hat{H}_0=\left(\begin{array}{cccc}
J_{z}+\epsilon_1+\epsilon_2&\Delta_2&\Delta_1&J_x-J_y\\
\Delta_2&\epsilon_1-\epsilon_2-J_{z}&J_{x}+ J_{y}&\Delta_1\\
\Delta_1&J_{x}+J_{y}&\epsilon_2-\epsilon_1-J_{z}&\Delta_2\\
J_{x}-J_{y}& \Delta_1&\Delta_2&-\epsilon_1-\epsilon_2+J_{z}\\
\end{array} \right) 
\end{equation}
In order to facilitate our analysis of the decohering effect of the environment, 
in what follows we assume the Ohmic nature of the dissipative reservoirs
described by the spectral function $S(\omega)=\int dte^{i\omega t}<h_{1,2}(t)h_{1,2}(0)>=
{\alpha\omega}\coth{\omega\over 2T}\Theta(\omega_c-\omega)$ of bandwidth $\omega_c$
(see the last part of this paper for a discussion of the relevance of this type
of noise for superconducting qubits).

The Ohmic nature of the noisy environment allows one to treat it
in the standard Bloch-Redfield (i.e., weak-coupling and Markovian) approximation.
In the basis spanned by the eigenstates of Eq.(\ref{5}),
the Bloch-Redfield equation for the reduced two-qubit density matrix reads
\cite{weiss}:
\begin{equation} 
\label{redfield} 
{\dot{\rho}_{nm}(t)=-i
\omega_{nm} \;\rho_{nm}(t)-\sum_{k,l}R_{nmkl}\;
\rho_{kl}(t)}, 
\end{equation} 
where $\omega_{nm}=(E_n-E_m)/\hbar$, $n,m=1,\dots, 4$ are the transition frequencies between the
eigenstates of Eq.(\ref{5}), and the relaxation tensor
\begin{equation} 
\label{r_tensor} 
R_{nmkl}=\delta_{lm}\sum_r \Lambda_{nrrk}+\delta_{nk}
\sum_r \Lambda^*_{lrrm}-\Lambda_{lmnk}-\Lambda^*_{knml}
\end{equation} 
is computed in terms of partial transition rates, whose real parts 
describe the effect of decoherence
\begin{eqnarray}
\label{9}
Re\{\Lambda_{lmnk}\}=\frac{1}{4\pi} S(\omega_{nk})\big[\sigma^z_{1,lm}\sigma^z_{1,nk}
+\sigma^z_{2,lm}\sigma^z_{2,nk}\big],
\end{eqnarray}
while $Im\{\Lambda_{lmnk}\}$ yield the Lamb shifts of the frequencies $\omega_{nk}$.

As a customary measure of the coupled qubits' performance we use gate purity
\begin{equation}
P(t)=\frac{1}{16}\sum_{j=1}^{16}Tr\{(\rho^j(t))^2\}
\end{equation} 
where the density matrix $\rho^j(t)$ is evaluated for a complete set of $16$
disentangled initial states: $\rho^j(0)=|\Psi_{in}^j><\Psi_{in}^j|$, 
$\Psi_{in}^j= |\Psi_a>_1\otimes |\Psi_b>_2,
(a,b=1,\dots,4$) which are composed of $|\Psi_1>=|\downarrow>$,
$|\Psi_2>=|\uparrow>$,$|\Psi_3>=(|\downarrow>+|\uparrow>)/{\sqrt
2}$, and $|\Psi_4>=(|\downarrow>+i|\uparrow>)/{\sqrt
2}$.

Below, we restrict our analysis to the Hamiltonians
that remain constant, $\hat{H}(t)=\hat{H}_0\theta(t)\theta(t_0-t)$, for the entire duration
$t_0$ the gate operation, and, therefore, commute at any times ($[\hat{H}(t),{\hat H}(t^\prime)]=0$).
We then demonstrate that within this class of quasi-stationary Hamiltonians, the
problem of constructing an optimized (coherence-wise) implementation 
of a given universal gate allows for a rather simple and physically transparent solution.

To that end, we utilize the mechanism of decoherence suppression proposed 
in Ref.\cite{ilya} where the decay rate of any initial state 
of an idling pair of coupled qubits ("quantum memory")
was shown to be related to the spectral properties of the Hamiltonian.

Namely, it was pointed out in Ref.\cite{ilya} that the relaxation-related component 
of the overall decoherence can be significantly reduced by
tuning the Hamiltonian parameters to the point in the parameter space where a pair of the lowest 
eigenvalues of the quasi-stationary Hamiltonian (\ref{5}) becomes degenerate
(see also the more recent Refs.\cite{others} for corroborating results).

The underlying (energy exchange-based) mechanism of the suppression of relaxation is based on
the observation that near a degeneracy point and at low temperatures the partial 
relaxation rates (\ref{9}) appear to be given by linear combinations of the transition frequencies
$\sum_{i\neq j}c_{ij}|\omega_{ij}|$ where the coefficients $c_{ij}$ are essentially 
independent of $\omega_{ij}$ \cite{ilya}. Therefore, the relaxation rates attain their minimum
values at those points in the parameter space where the largest possible number of the 
transition frequencies vanish due to degeneracy.
By contrast, a contribution from the other, pure dephasing, processes
is generally unavoidable and can only be suppressed by lowering the temperature of the reservoirs.

In order to further illustrate the above point, 
in Fig.1 we plot the absolute value of the purity decay rate $|dP/dt|$
(which, when computed in the Markovian approximation,
remains approximately constant at times shorter than 
the decoherence time) as a function of the components of the inter-qubit coupling 
$J_y$ and $J_z$, while keeping the single-qubit fields ${\vec B}_i$ constant and 
the length of the vector $J={\sqrt {J_x^2+J_y^2+J_z^2}}$ fixed. On physical grounds, the latter constraint
is justified by the fact that
in any realistic qubit system an unlimited 
increase of $J$ would eventually result in an unwanted leakage from the designated
two-qubit Hilbert susbspace. 

The color plot of Fig.1 demonstrates that $|dP/dt|$ does attain its absolute 
minimum at the points characterized by the incidence of double degeneracy
between the eigenvalues
of Eq.(3) (see Eq.(15) below).
Furthermore, albeit being less effective than its double counterpart, 
the onset of even a single degeneracy between the two lowest
eigenvalues appears to provide a relative improvement, as compared to the 
generic (non-degenerate) situation. Even in this case, the two-dimensional degenerate ground state
gets protected by the energy gap separating it from the rest of the spectrum
that gives rise to the exponential suppression of relaxation at low temperatures.
Considering the challenge of a high-precision tuning of the qubits' couplings in any realistic setup, 
the possibility of improving the gate quality with such a more relaxed constraint on the Hamiltonian parameters
can be of a particular interest for practical realizations of robust quantum gates.
 
Having identified the conditions required for a suppression of decoherence, 
we can now attempt to satisfy them in the cases of some popular universal gates. 
For a starter, we impose the less stringent condition of 
a single degeneracy between the lowest pair of energy
levels, while looking for a one-step implementation
of the standard CNOT gate $\hat{X}_{CNOT}=
diag([1\;0\;0\;0][0\;1\;0\;0][0\;0\;0\;1][0\;0\;1\;0])$.

In the noiseless case, this goal can be accomplished by directly computing
the matrix-valued logarithm of the unitary matrix 
$\hat{X}_{CNOT}$. It turns out, however, that the latter features a high degree of ambiguity
\begin{equation}
\label{10} 
\hat{H}_{CNOT} =i/t_{0} \ln{(\hat{X}_{CNOT})}=C(A+B)C^{-1} 
\end{equation}
Here
$t_0$ is the total duration of the gate operation
\begin{equation}
A=\left(\begin{array}{cccc} 0&0&0&0\\ 0&0&0&0\\ 0&0&-\pi/2&\pi/2\\
0&0&\pi/2&-\pi/2\\ \end{array} \right)+\phi_0 \hat{E},
\end{equation}
$\hat{E}$ is the unit matrix accounting for the possible "mismatch" phase
$\phi_0$, while the matrices
\begin{equation}
B=2\pi n_1 \left(\begin{array}{cccc} 1&1&0&0\\1&1&0&0\\0&0&1&1\\0&0&1&1\\
 \end{array} \right)+
2\pi n_2\left(\begin{array}{cccc} 0&0&1&1\\0&0&1&1\\1&1&0&0\\1&1&0&0\\
 \end{array} \right)+
2\pi  n_3 \hat{E},
\end{equation}
(observe that $[A,B]=0$) and
\begin{equation}
C=\left(\begin{array}{cc} e^{i\vec{\phi}\vec{\sigma}} &0\\
0&e^{i\phi_1\sigma_x} \end{array} \right) 
\end{equation} 
manifest the multi-valuedness of the matrix logarithm of the CNOT gate
which can be obtained by exponentiating Eq.(10) for an arbitrary choice of the 
integer $(n_i)$ as well as continuous $(\vec \phi$, $\phi_1)$ parameters in Eqs.(12) and (13), respectively. 
The existence of such an extensive invariant subspace within the equivalence 
class of CNOT (see below) is a special property of this particular gate.

A straightforward analysis reveals that the CNOT gate can indeed be reproduced 
up to a global phase $\phi_0=-\pi/4$
(so that $\hat{X}_{CNOT}=\exp(-i\pi/4-it_{0}\hat{H}_0)$) 
under the condition of a single spectral degeneracy 
($E_1=E_2=-0.875, E_3=0.625, E_4=1.125$) with the following choice of the parameters in Eq.(\ref{10}):
$\Delta_1=J_x=J_y=0, \Delta_2=1.5, \epsilon_1=-0.25, \epsilon_2=J_z=-0.66$  
(hereafter all the energies are in units of $1/t_0$).
 
In Fig.2 we contrast the resulting optimized gate
against the standard CNOT protocol consisting of a series of non-overlapping pulses 
(see, e.g., \cite{whaley})
\begin{eqnarray}
\hat{X}_{CNOT}=
\exp\big(-i\frac{\pi}{2}\frac{\sigma^x_2+\sigma^z_2}{\sqrt{2}}
\big)\exp\big(-i\frac{\pi}{2}\sigma^z_1\big)\nonumber\\
\exp\big(-i\frac{\pi}{2}
\sigma^z_2\big) 
\exp\big(-i\frac{\pi}{4}\sigma^z_1\sigma^z_2\big)
\exp\big(-i\frac{\pi}{2}\frac{\sigma^x_2+\sigma^z_2}{\sqrt{2}}\big)
\end{eqnarray}
 
In order to compare the quality of the two protocols we solve the corresponding 
Bloch-Redfield equations (4) in the presence of weak ($\alpha<<1$) noise in both cases. As a result, we find 
that the one-step implementation,
where both the two- and one-qubit operations are carried out simultaneously,
takes about $15\%$ of the time of
and appears to be approximately $10$ times better (in terms of the gate purity) 
than the standard protocol (12).

We mention, in passing, that one can also carry out
a one-step implementation of other commonly used (e.g., RNOT and fast Fourier transform) gates
by employing the Hamiltonians with single spectral degeneracies, thus significantly improving the
gates' performances.
 As the next step, we attempt to impose the double degeneracy condition that, if at all attainable, 
is expected to result in an even better performance. It is, however, conceivable that this more
stringent condition may not necessarily allow one to construct an arbitrary two-qubit gate.

Nonetheless, even if this turns out to be the case, one 
can still elect to settle for a more readily achievable goal of 
constructing a gate $X^\prime$ which is equivalent
(i.e. $X^\prime=U_1\otimes U_2XU_1^\prime\otimes U^\prime_2$ where $U_{1,2}$ and $U^\prime_{1,2}$
are single-qubit rotations), albeit not necessarily identical, to a target gate $X$.
It should be noted, though, that in practice such a substitute solution might only be acceptable
if all the one-qubit rotations can be performed sufficiently quickly, thus contributing negligibly 
towards the overall loss of coherence. 

Formally, the equivalence between a pair of gates is established on the basis of 
a coincidence of their Makhlin's invariants (see, e.g., \cite{whaley}) 
\begin{eqnarray}
G_1=\frac{tr^2[m(\hat{X})]}{16 Det[\hat{X}]},
~~~G_2=\frac{tr^2[m(\hat{X})]-tr[m^2(\hat{X})]}{4 Det[\hat{X}]}
\end{eqnarray}
where $m(\hat{X})=\hat{X}^T_B \hat{X}_B$, $\hat{X}_B=\hat{Q}^\dagger \hat{X} \hat{Q}$, and 
$\hat{Q}=diag([1\;0\;0\;i][0\;i\;1\;0][0\;i\;-1\;0][1\;0\;0\;-i])$.

A straightforward analysis shows that in the presence of a double spectral degeneracy
$G_1$ always turns out to be a real number, which fact severely restricts
the set of gates that can be constructed this way.
One example of a gate incompatible with the double degeneracy condition is provided by  
the $\sqrt{SWAP}$ gate with $G_1=-i/4$. 
By contrast, the CNOT equivalence class (albeit not the original CNOT gate itself!) with the 
invariants $G_1=0$, $G_2=1$ can be readily attained under the condition of double degeneracy.

Our theoretical findings can be of an immediate relevance to such a viable
candidate to the role of robust two-qubit gates as a pair of
the charge/flux Josephson junction qubits.
Therefore, in what follows we specifically discuss this particular realization 
and focus on its most coherence-friendly regime (dubbed "quantronium" in Ref.\cite{devoret})
where the individual qubits are tuned to their optimal points $\epsilon_{1,2}=0$ 
and coupled together both, capacitively \cite{devoret} and inductively \cite{mooji}.

At the optimal point, the eigenvalues of Eq.(3) assume a particularly simple form
\begin{eqnarray}
E_{1,2}=J_x\mp{\sqrt{(\Delta_1+\Delta_2)^2+(J_y-J_z)^2}},\nonumber\\
E_{3,4}=-J_x\pm{\sqrt{(\Delta_1-\Delta_2)^2+(J_y+J_z)^2}}
\end{eqnarray}
Considering, for the sake of simplicity, the case of two identical qubits
($\Delta_{1,2}=\Delta$), one can readily show that
the double degeneracy condition ($E_1=E_4, E_2=E_3$) is achieved at
\begin{equation}
J_x=0, ~~~~J_yJ_z=\Delta^2
\end{equation}
Thus, in order to implement a transformation that belongs to the CNOT equivalence class under
the double degeneracy condition it suffices to use a single rectangular interaction pulse 
with the parameters
\begin{equation}
{\vec J}_{CNOT}=(0,{1\over {\sqrt 2}}({\sqrt{(J^2+\Delta^2}}-{\sqrt{J^2-\Delta^2}}), 
{1\over {\sqrt 2}}({\sqrt{(J^2+\Delta^2}}-{\sqrt{J^2-\Delta^2}}))
\end{equation}
From the Bloch-Redfield equations we find that corresponding
protocol outperforms 
the CNOT-equivalent gate constructed in Ref.\cite{romero} with the use
of the Heisenberg inter-qubit coupling, as far as  
both the duration ($\approx 2.3$ shorter) and robustness ($\approx 25$ times
lower purity decay rate) are concerned, thereby resulting 
in $\approx 60$ times lower overall purity loss $1-P(t_0)$.

For a large part, the uqibuitous CNOT gate owes its high popularity to the fact that 
a generic two-qubit gate requires no more than three CNOT applications complemented by
local single-qubit rotations \cite{3not}. Recently, 
the authors of Ref.\cite{whaley} put forward an alternative (dubbed as the "B") gate 
with the invariants $G_1=G_2=0$.
Unlike CNOT, however, the B-gate only needs to be used 
twice in order to implement an arbitrary two-qubit gate (up to single qubit rotations).

With an eye on the possibility of a further optimization, 
we find that our approach which takes a full advantage of the double degeneracy condition 
offers a superior realization of the B-gate universality class, too. 
Namely, by choosing $\Delta_{1,2}=1., \epsilon_{1,2}=J_x=0, J_y=0.58, J_z=1.71$ (in units of $1/t_0$) 
we arrive at the one-step implementation of the equivalent of the B-gate
that takes only $\approx 56\%$ of the duration of the protocol of Ref.\cite{whaley}
and results in a much improved  ($\approx 4.5$ times slower) purity decay rate, thus decreasing the  
purity loss $1-P(t_0)$ by a factor of $\approx 8$.

Finally, we apply our theoretical analysis to the two-qubit Josephson junction 
setups of Ref.\cite{pashkin} with the parameters $\Delta\sim 10$ GHz and $J\sim 20$ GHz.
To that end, we first extract the pertinent value $\alpha\sim 0.01$ of the dissipative coupling from
the noise spectral function. The latter is simply proportional to the single-qubit relaxation rate,
$1/T_1={\pi\over 2}S(\Delta)$, which was directly measured right at the optimal point and found to be of order
$1/T_1\approx 0.1$GHz (see the last reference in Ref.\cite{pashkin}). 

With these data at hand, we estimate the purity decay which would occur in the course 
of implementing the above equivalent of the B-gate as $1-P_{B}(t_0)\approx 0.03$.
However, this estimate would only hold for temperatures above $\sim 0.1$K (albeit smaller than $\Delta$),
for at still lower temperatures decoherence was found to be dominated by the non-Ohmic (most likely $1/f$)
noise which can not be treated the same way as the Ohmic one within the Bloch-Redfield approach.

Correspondingly, in the case of the before mentioned equivalent of the CNOT gate one would obtain  
$1-P_{CNOT}(t_0)\approx 0.15$ which compares favorably to the result
$1-P_{CNOT}(t_0)\approx 0.27$ obtained in the second reference of Ref.\cite{pashkin}.
However, our theoretical prediction still falls too short of the maximally acceptable error rate
$\sim 10^{-4}$ which can be tackled with the existing error-correction codes.
A further improvement would then require a coherence-conscience engineering of the environment,
with the aim at significantly reducing the parameter $\alpha$.  

It should be noted, however, that the authors of Ref.\cite{pashkin} 
attributed the loss of purity measured in that work 
solely to the errors in controlling the shape of the operation pulses (finite rise/fall time).
For comparison, the sensitivity of our optimized protocols, in the presence of the intraction with reservoirs,
 with respect to variations $\delta x_i$ of 
the control parameters appears to be reduced, thanks to the quadratic (as opposed to a generic linear in the case
of not coherence-wise optimization) dependence
upon such variations near the minima of the deviation functional (1).

For example, we find that in order to keep $1-P(t_0)$ under the $10^{-4}$ threshold
the tolerance limit of (de)tuning of the optimized CNOT and B-gates' parameters has to be better than
$0.3\%$ of their values. 

Before concluding, it is worth mentioning that 
our approach differs from the previous proposals for constructing "supercoherent" qubits
which, albeit being capable of providing an exponential suppression of decoherence,
require at least four physical qubits which are governed by the Hamiltonian 
that conserves total spin \cite{whaley2}.
Apart from a larger number of physical qubits required to encode a logical one,
the condition of spin rotational invariance of the Hamiltonian is unlikely 
to be fulfilled in any realistic solid-state qubits where (contrary to liquid-state NMR designs) 
the instantaneous values 
of the single- and two-qubit terms in Eq.(2) are often related, thus 
resulting in the local terms (${\vec B}\neq 0$) which break the spin-rotational invariance.

To summarize, we present a systematic approach to constructing simplified (one-step) and optimized
(highly robust) implementations of various two-qubit universal gates. 
The high stability against decoherence is achieved, thanks to the choice of the Hamiltonian parameters 
that provide for the degeneracies
in the instantaneous energy spectrum, 
thereby resulting in the suppression of relaxation processes. 
Finally, we anticipate that the general nature of the decoherence suppression mechanism
exploited in this work should make it possible to further generalize our results 
to the sequences of two-qubit as well as irreducible multi-qubit gates.

This research was supported by ARO under Contract DAAD19-02-1-0049 and by NSF under Grant DMR-0349881.

\begin{figure}
\begin{center}
\includegraphics[height=0.4\textwidth,angle=0]{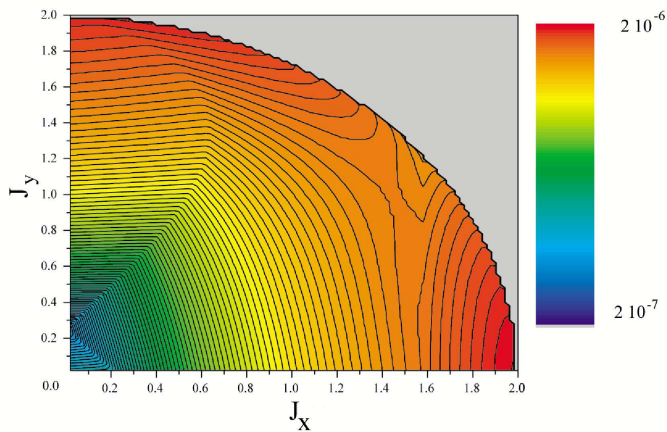} \vspace{0.3cm}
\caption {Gate purity decay rate $|dP/dt|$ as a function of interaction coefficients $J_x$ and $J_y$}
\end{center}
\end{figure}

\begin{figure}
\begin{center}
\includegraphics[height=0.3\textwidth,angle=0]{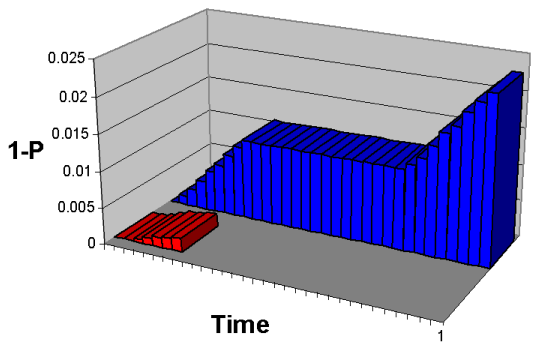} \vspace{0.3cm}
\caption {Comparison between the standard (five-step) implementation 
and the optimized one-step implementations of the CNOT gate.}

\end{center}
\end{figure}

\end{document}